\documentclass[12pt]{article}
\usepackage{graphicx}
\usepackage{amsmath}
\usepackage{amssymb}
\usepackage{enumitem}
\usepackage{latexsym}
\begin{document}
\vskip 2cm

\def\warning#1{\begin{center}
\framebox{\parbox{0.8\columnwidth}{\large\bf #1}}
\end{center}}

\begin{center}

{\sf {\bf Dimensional reduction of four-dimensional topologically massive gauge theory}}

\vskip 2.5 cm

{\sf{ \bf R. Kumar\footnote{E-mail: rohit.kumar@bose.res.in; raviphynuc@gmail.com} and Amitabha Lahiri\footnote{E-mail: amitabha@bose.res.in}}}\\
\vskip .1cm
{\it S. N. Bose National Centre for Basic Sciences,\\
Block JD, Sector III, Salt Lake, Kolkata$-$700098, India}\\
\end{center}

\vskip 2.5 cm

\noindent

\noindent
{\bf Abstract:} We study topologically massive $(B \wedge F)$ Abelian and non-Abelian gauge theories in four dimensions, and reduce them to three dimensions by assuming that the fields do not propagate in one of the spatial directions. For certain gauge choices the  reduced theories  are the Jackiw-Pi models of massive gauge fields.

\vskip 1cm
\noindent
{ PACS:} 11.15.-q, 03.70.+k, 11.10Kk, 12.90.+b\\

\noindent
{\it Keywords}: $(B\wedge F)$ model;  Jackiw-Pi model; dimensional reduction\\

\newpage

\section{Introduction}
Recently, there has been a great deal of theoretical interest in the study of massive Abelian and non-Abelian gauge theories in four dimensions, in which gauge invariance and mass of the gauge bosons co-exist in a particular fashion. Several such different massive gauge invariant theories are known in four spacetime dimensions~\cite{aur,pk,trg,a1,ru, a2}. One such model is the $(3+1)$-dimensional topologically massive gauge theory where a 1-form gauge field is coupled to a 2-form antisymmetric gauge potential through a topological $B \wedge F$ term. The 1-form gauge bosons acquire mass through this topological term.

These theories have been studied from many different angles such as Hamiltonian and constraint analysis~\cite{a3,a4}, BRST quantization~\cite{a5,r1,r2,r3}, geometric construction~\cite{mieg,lee} and the phenomenological point of view~\cite{a6}. The perturbative renormalizability of the dynamical 2-form non-Abelian theory is shown algebraically in~\cite{a2} by using the Zinn-Justin equation in the context of BRST formalism where an additional compensating auxiliary vector field is introduced in order to bypass the well-known no-go theorem~\cite{hen}.

On the other hand, massive gauge theories are also known in lower dimensions~\cite{j1,j2,j3}. One of these models is the $(2+1)$-dimensional Jackiw-Pi model~\cite{j3}. This model has been studied from many different point of views such as Hamiltonian and correct constraints analysis~\cite{day}, the establishment of Slavnov-Taylor identities and BRST symmetries~\cite{cim}. Furthermore, the Jackiw-Pi model is known to be ultraviolet finite and renormalizable~\cite{cim}. The off-shell nilpotent and absolutely anticommuting (anti-)BRST transformations for this model, as well as the geometrical interpretation of these nilpotent symmetries have been discussed within the framework of Bonora-Tonin superfield formalism~\cite{r4,r5}.

Even though the four dimensional $(B \wedge F)$ model appears to be different from the three-dimensional Jackiw-Pi model at first glance, there  are also some obvious similarities between them~\cite{r5}. Both models are gauge invariant, and the gauge field acquires mass through a topological term without taking any help of Higgs mechanism. In addition, they respect two independent sets of gauge symmetries, the usual SU(N) gauge symmetry as well as a non-compact gauge symmetry. As a consequence, one can naturally ask a question. Can these two different models of mass generation of the gauge field, in different spacetime dimensions, be related to each other? In  our present investigation, we pursue this question and establish a possible connection between these two different models with the help of dimensional reduction technique.

Dimensional reduction is a standard procedure in Kaluza-Klein theories and supergravity~\cite{sch,c1,c2,sal}. In our present endeavor, we use a different dimensional reduction procedure where the fields in higher dimensions are assumed to be independent of the extra dimension~\cite{siv3}. This technique has been applied to various models defined in even as well as odd dimensions of spacetime (see,  e.g.~\cite{siv3,net1}) and the reduced models show quite interesting and surprising features. For instance, Chern-Simon theory does not respect parity and time-reversal symmetries.  However, it has been shown in~\cite{siv3} that, using the dimensional reduction procedure, one obtains the bosonized Schwinger model~\cite{sch1,sch2}, which respects parity as well as time reversal symmetries provided the intrinsic parity and time reversal properties of the scalar field are chosen opposite to those of the vector field.

The contents of our present paper are organized as follows. In Sec.~\ref{jpm}, we briefly discuss the Jackiw-Pi model and associated local gauge symmetries. In Sec.~\ref{topomass} we start with the topologically massive Abelian theory in four dimensions and reduce it to  three dimensions via a dimensional reduction procedure where we assume that the all the fields are independent to the third spatial  direction. Sec.~\ref{nonab} deals with the dimensional reduction of the non-Abelian version of the four dimensional model to three dimensions. 

{\sl Conventions and notation:} We adopt conventions and notation such that the four-dimensional Minkowski metric is $\eta_{\bar \mu \bar \nu} =$ diag $(+ 1, - 1, -1, -1)$ and the three-dimensional metric is $\eta_{\mu\nu} =$ diag $(+1, - 1, -1)$. Components of four-dimensional objects will be denoted by Greek indices with a bar on top, $\bar \mu, \bar \nu, \bar \eta, \cdots = 0, 1, 2, 3\,,$ whereas compoennts of three-dimensional objects will be denoted by Greek indices without a bar,  $\mu, \nu, \eta, \cdots = 0, 1, 2\,.$ The Levi-Civita tensors $\varepsilon^{\bar \mu \bar \nu \bar \eta \bar \kappa}$ and  $\varepsilon^{\mu\nu\eta}$ are chosen such that $\varepsilon^{0123} = \varepsilon^{012} = +1 = - \varepsilon_{0123} = + \varepsilon_{012},$ etc.
The $SU(N)$ generators $T^a$ (with $a, b, c,... = N^2 - 1$) satisfy the commutation relation $[T^a,\, T^b] = i f^{abc}\, T^c$ where the structure constants $f^{abc}$ are chosen to be totally antisymmetric in the indices. In the $SU(N)$ Lie algebra,  the dot and cross products between two  vectors $P$ and $Q$ are defined as $P \cdot Q = P^a Q^a,\; P \times Q = f^{abc}\, P^a Q^b T^c$.

\section{Jackiw-Pi model}\label{jpm}
We begin with the $(2+1)$-dimensional non-Abelian Jackiw-Pi  model \cite{j3}. The classical Lagrangian density incorporates the 
topological mass parameter $m$ through a mixed Chern-Simons term given by 
\begin{eqnarray}
{\cal L}_{(JP)} &=& - \frac{1}{4}\, F_{\mu \nu} \cdot F^{ \mu \nu} 
- \frac{1}{4}\, \big[G_{\mu\nu} + g (F_{\mu\nu} \times \rho) \big] \cdot \big[G^{\mu\nu} + g (F^{\mu\nu} \times \rho)\big]\nonumber\\
&&+ \frac{m}{2}\,\varepsilon^{\mu\nu\eta}\,F_{\mu\nu}\cdot \phi_\eta, 
\label{JP}
\end{eqnarray}
where $F_{\mu\nu} \equiv F^a_{\mu\nu}\,T^a = \partial_\mu A_\nu - \partial_\nu A_\mu + g\,(A_\mu \times A_\nu)$ is the usual 2-form field strength tensor corresponding to the 1-form gauge field $A_\mu \equiv A^a_\mu \, T^a$. The compensated curvature tensor for the other 1-form gauge field $\phi_\mu \equiv \phi^a_\mu \, T^a$ is defined with the help of an auxiliary scalar field $\rho \equiv \rho^a \,T^a$ as given by
\begin{eqnarray}
\big[G^a_{\mu\nu} + g\,(F_{\mu\nu} \times \rho)^a]\, T^a 
&=& D_\mu \phi_\nu - D_\nu \phi_\mu + g\,(F_{\mu\nu} \times \rho), 
\end{eqnarray}
where $D_\mu \Omega = \partial_\mu \Omega + g \,(A_\mu \times \Omega)$ defines the covariant derivative for any generic field $\Omega \equiv \Omega^a\, T^a$ in the adjoint representation.
The fields $A_\mu\,,\phi_\mu\,,$ and coupling constant $g$ all have mass dimension equal to $[M]^{\frac{1}{2}}$ whereas the auxiliary field $\rho$ carries mass dimension $[M]^{-\frac{1}{2}}$. The constant $m$ has the dimension of mass, and the gauge field $A_\mu$ acquires a mass $m$ in this model without breaking the gauge symmetry. The vector fields $A_\mu$ and $\phi_\mu$ may be chosen to have opposite parities which makes this model parity invariant~\cite{j3}, as opposed to the usual Chern-Simons theory in which the term $\frac{m}{2}\,\varepsilon^{\mu\nu\eta}\,F_{\mu\nu}\cdot A_\eta$ breaks the parity symmetry \cite{j1,j2}.

The Lagrangian density respects the two independent sets of gauge transformations. The first is the usual SU(N) gauge transformation, and the second is a non-compact, St\"uckelberg-type transformation~\cite{j3,cim}
\begin{eqnarray}
&&\delta_1\, A_\mu = D_\mu \Lambda,  \quad \delta_1\, \phi_\mu = g (\phi_\mu \times \Lambda), 
\quad \delta_1\, \rho = g (\rho \times \Lambda), \nonumber\\
&&\nonumber\\
&& \delta_2 \,\phi_\mu = D_\mu \Sigma, \quad \delta_2\, \rho = - \Sigma, \quad \delta_2 \,A_\mu = 0, 
\label{JP_gt}
\end{eqnarray}
where $\Lambda \equiv \Lambda^a\,T^a$ and $\Sigma \equiv \Sigma^a\,T^a$ are the $SU(N)$ Lie algebra valued local gauge  parameters. It is straightforward to check that  the Lagrangian density remains invariant under the $SU(N)$ gauge transformations while the other set of gauge transformations changes the  Lagrangian density by a total spacetime derivative,
\begin{eqnarray}
\delta_1\,{\cal L}_{(JP)} = 0, \qquad 
\delta_2\,{\cal L}_{(JP)} = \partial_\mu \left[\frac{m}{2}\,\varepsilon^{\mu\nu\eta}\,F_{\mu\nu}\cdot \Sigma\right]. 
\end{eqnarray}
As a consequence, the action remains invariant under both symmetries for configurations of fields which vanish sufficiently rapidly at infinity.

The Abelian version of the Jackiw-Pi model may be written in the form
\begin{eqnarray}
\tilde {\cal L}_{(JP)} &=& - \frac{1}{4}\, F_{\mu \nu} \,F^{ \mu \nu} - \frac{1}{4}\, G_{\mu\nu}\,G^{\mu\nu} 
+ \frac{m}{2}\,\varepsilon^{\mu\nu\eta}\,F_{\mu\nu}\,\phi_\eta, 
\label{AJP}
\end{eqnarray}
where $F_{\mu\nu} = \partial_\mu A_\nu -\partial_\nu A_\mu$ and $G_{\mu\nu} = \partial_\mu \phi_\nu -\partial_\nu \phi_\mu$ are the Abelian 
field strength tensors for the fields $A_\mu$ and $\phi_\mu$, respectively. 
One can check that this Lagrangian density respects the following local gauge transformations:
\begin{eqnarray}
&& \delta_1\, A_\mu = \partial_\mu \Lambda, \qquad \delta_1\, \phi_\mu = 0,\nonumber\\
&& \delta_2\, \phi_\mu = \partial_\mu \Sigma, \qquad \delta_2\, A_\mu = 0, 
\label{AJP_gt}
\end{eqnarray}
where $\Lambda$ and $\Sigma$ are now the local gauge transformation parameters.

\section{Topologically massive Abelian theory}\label{topomass}
In this section, we discuss the $(3+1)$-dimensional topologically massive Abelian gauge theory and reduce it to a $(2+1)$-dimensional theory. The Lagrangian density of this theory, which contains an Abelian gauge field and an Abelian 2-form field, is given by~\cite{a1} 
\begin{eqnarray}
\tilde {\cal L}_{(BF)} = - \frac{1}{4}\, F_{\bar \mu \bar \nu} \,F^{\bar \mu \bar\nu} 
+ \frac{1}{12}\, H_{\bar \mu \bar \nu \bar \eta}\, H^{\bar  \mu \bar \nu \bar \eta} 
+ \frac{m}{4}\, \varepsilon^{\bar \mu \bar \nu \bar \eta \bar \kappa}\,B_{\bar \mu \bar \nu}\, F_{\bar \eta\bar \kappa},
\label{L_BF}
\end{eqnarray} 
where $F_{\bar \mu \bar \nu} = \partial_{\bar \mu} A_{\bar \nu} - \partial_{\bar \nu} A_{\bar \mu}$ is the Abelian field strength 
tensor corresponding to the 1-form gauge field $A_{\bar \mu}$, the field strength for the antisymmetric gauge potential $B_{\bar \mu \bar\nu}$ is defined as $H_{\bar \mu \bar \nu \bar \eta} = \partial_{\bar \mu} B_{\bar \nu \bar \eta} + \partial_{\bar \nu} B_{\bar \eta \bar \mu} + \partial_{\bar \eta} B_{\bar \mu \bar \nu}$ and $m$ represents the topological mass parameter which has mass dimensions $[M]$. In four dimensions, all the fields carry the same mass dimension equal to $[M]$.

The Lagrangian density of Eq.~(\ref{L_BF}) respects two independent local gauge symmetry transformations, 
\begin{eqnarray}
&&\delta_1\, A_{\bar \mu} = \partial_{\bar \mu} \Lambda, \qquad \delta_1\, B_{\bar \mu \bar \nu} = 0\nonumber\\
&&\delta_2\, B_{\bar \mu \bar \nu} = \partial_{\bar \mu} \Sigma_{\bar \nu} - \partial_{\bar \nu} \Sigma_{\bar \mu},
\qquad \delta_2\, A_{\bar \mu} = 0,
\label{Ab_gt}
\end{eqnarray} 
where $\Lambda$ and $\Sigma_{\bar \mu}$ are the scalar and vector gauge transformation parameters, respectively. Like the Jackiw-Pi model in three dimensions, this model generates mass for the gauge field $A_\mu$ in four dimensions, without spoiling the gauge symmetries.

In order to reduce this model to three dimensions, we will assume that all the fields defined in four dimensions are independent of the third spatial coordinate $x^3$. Then we split the fields in the following fashion~\cite{siv3,net1},
\begin{eqnarray}
A_{\bar \mu} &\equiv& (A_\mu,\, A_3) \equiv (A_\mu,\, \varphi),\nonumber\\
B_{\bar \mu \bar \nu} &\equiv& (B_{\mu\nu},\, B_{\mu 3}) \equiv (B_{\mu\nu},\, \phi_\mu), \qquad (\mu, \nu,...= 0,1,2).
\end{eqnarray}
Setting the $x^3$ derivatives of all fields to zero,  we obtain the following reduced  effective Lagrangian density  
\begin{eqnarray}
\tilde {\cal L}_{(BF)} &=& - \frac{1}{4}\, F_{\mu \nu} \,F^{ \mu \nu} 
- \frac{1}{4}\, G_{\mu\nu}\,G^{\mu\nu} + \frac{m}{2}\,\varepsilon^{\mu\nu\eta}\,F_{\mu\nu}\,\phi_\eta
+ \frac{1}{12}\, H_{\mu \nu  \eta}\, H^{\mu \nu \eta} \nonumber\\
&&- \frac{m}{3!}\, \varepsilon^{\mu\nu\eta}\,H_{\mu\nu\eta}\,\varphi 
+ \frac{1}{2}\,\partial_\mu \varphi\, \partial^\mu \varphi + \frac{m}{2}\,\partial_\mu\big[\varepsilon^{\mu\nu\eta} B_{\nu\eta}\, \varphi\big],
\end{eqnarray} 
modulo a total derivative term, where we have defined $G_{\mu\nu} = \partial_\mu \phi_\nu - \partial_\nu \phi_\nu$.
This Lagrangian density can further be simplified by noting that the Hodge dual of the 3-form $H_{\mu\nu\eta}$ in three dimensions is a scalar, $f = \frac{1}{3!}\,\varepsilon^{\mu\nu\eta}\, H_{\mu\nu\eta}$. This allows us to write
\begin{eqnarray}
\tilde {\cal L}_{(BF)} &=& - \frac{1}{4}\, F_{\mu \nu} \,F^{ \mu \nu} 
- \frac{1}{4}\, G_{\mu\nu}\,G^{\mu\nu} + \frac{m}{2}\,\varepsilon^{\mu\nu\eta}\,F_{\mu\nu}\,\phi_\eta 
 + \frac{1}{2}\,\partial_\mu \varphi\, \partial^\mu \varphi + \frac{1}{2}\,f^2 - m\,f\,\varphi\nonumber\\
&\equiv& - \frac{1}{4}\, F_{\mu \nu} \,F^{ \mu \nu} - \frac{1}{4}\, G_{\mu\nu}\,G^{\mu\nu} 
+ \frac{m}{2}\,\varepsilon^{\mu\nu\eta}\,F_{\mu\nu}\,\phi_\eta + \frac{1}{2}\,\partial_\mu \varphi\, \partial^\mu \varphi
- \frac{1}{2}\,m^2 \varphi^2,
\label{dr_ab_lag}
\end{eqnarray}
where in the last step we have used the equation of motion $f  = m\, \varphi$  for the field $f$. An alternative way of seeing this is by completing a square in the path integral. The Lagrangian density of Eq.~(\ref{dr_ab_lag}) coincides with that of the Abelian Jackiw-Pi model~\cite{j3}, 
plus that of a real massive scalar $\varphi\,,$ decoupled from each other.

Furthermore, the gauge transformations of Eq.~(\ref{Ab_gt}) can be written after dimensional reduction as
\begin{eqnarray}
&& \begin{cases}
\delta_1 A_\mu = \partial_\mu \Lambda,\qquad \delta_1 \varphi = 0,\\
\delta_1 B_{\mu\nu} = 0, \qquad \delta_1 \phi_\mu = 0,\\
 \end{cases} 
\nonumber\\
&&
\begin{cases}
\delta_2 B_{\mu \nu} = \partial_\mu \Sigma_\nu - \partial_\nu \Sigma_\mu, \qquad
\delta_2 \phi_\mu = \partial_\mu \Sigma,
 \nonumber\\
\delta_2 A_\mu = 0, \qquad \delta_2 \varphi = 0,
\end{cases} 
\end{eqnarray}
where we have defined $\Sigma \equiv \Sigma_3$. The dimensionally reduced action respects the above gauge symmetries, since one can explicitly check that 
\begin{eqnarray}
\delta_1 \tilde {\cal L}_{(BF)} = 0, \qquad 
\delta_2 \tilde {\cal L}_{(BF)} = \partial_\mu\Big[\frac{m}{2}\, \varepsilon^{\mu\nu\eta}\, \Sigma\, F_{\nu\eta}\Big].
\end{eqnarray}
Clearly, if we consider the four-dimensional theory in the axial gauge $A_3 = 0$ or equivalently $\varphi = 0$, we find the Abelian Jackiw-Pi model in three dimensions along with its associated local gauge transformations. Thus, the dimensional reduction procedure establishes a connection between the two models.

Before we wrap up this section, we note that if the gauge field is minimally coupled to fermions in four dimensions, we find a scalar with a Yukawa-type coupling in the reduced theory. However, if we select the axial gauge in four dimensions, $\varphi$ vanishes in the three-dimensional theory, and we are left with only the Jackiw-Pi model.

\section{Topologically massive non-Abelian theory}\label{nonab}
The non-Abelian generalization of the topological mass generation mechanism is not a straightforward task because of a no-go theorem~\cite{hen}. However, one can avoid this theorem by introducing a compensating auxiliary field in a particular fashion. Thus, the non-Abelian generalization of Eq.~(\ref{L_BF}) is given by the following Lagrangian density~\cite{a2}
\begin{eqnarray}
{\cal L}_{(BF)} = - \frac{1}{4}\, F_{\bar \mu \bar \nu} \cdot F^{\bar \mu \bar\nu} 
+ \frac{1}{12}\, H_{\bar \mu \bar \nu \bar \eta}\cdot H^{\bar  \mu \bar \nu \bar \eta} 
+ \frac{m}{4}\, \varepsilon^{\bar \mu \bar \nu \bar \eta \bar \kappa}\,B_{\bar \mu \bar \nu}\cdot F_{\bar \eta\bar \kappa},
\label{L_NA}
\end{eqnarray} 
where $F^a_{\bar\mu \bar \nu}\,T^a \equiv F_{\bar \mu \bar \nu} = \partial_{\bar \mu} A_{\bar \nu} 
- \partial_{\bar \nu} A_{\bar \mu} + g\,(A_{\bar \mu} \times A_{\bar \nu})$ is the field strength 
tensor for the non-Abelian gauge field $A^a_{\bar \mu}\,T^a$, with $T^a$ being the $SU(N)$ generators, and $g$ is a dimensionless coupling constant. The totally antisymmetric compensated curvature tensor $H_{\bar \mu \bar \nu \bar \eta}$  for the gauge field $B^a_{\bar \mu \bar \nu}\,T^a$ is defined as 
\begin{eqnarray} 
H_{\bar \mu \bar \nu \bar \eta} &=& D_{[\bar \mu} B_{\bar \nu \bar \eta]}
+ g\,(F_{[\bar\mu \bar \nu} \times K_{\bar \eta]})\nonumber\\
&=& D_{\bar \mu} B_{\bar \nu \bar \eta} + D_{\bar \nu} B_{\bar \eta \bar \mu} 
+ D_{\bar \eta} B_{\bar \mu \bar \nu} \nonumber\\
&&+ g\,(F_{\bar\mu \bar \nu} \times K_{\bar \eta}) + g\,(F_{\bar\nu \bar \eta} \times K_{\bar \mu})  
+ g\,(F_{\bar\eta \bar \mu} \times K_{\bar \nu}),
\end{eqnarray}
where $K_{\bar \mu}$ is an auxiliary vector field. The Lagrangian density of Eq.~(\ref{L_NA}) remains invariant under the usual SU(N) gauge transformations 
\begin{eqnarray}
\delta_1\, A_{\bar \mu} = D_{\bar \mu} \Lambda, \qquad \delta_1\, B_{\bar \mu \bar \nu} 
= g (B_{\bar \mu \bar \nu} \times \Lambda), \qquad \delta_1\, K_{\bar \mu} = g (K_{\bar \mu} \times \Lambda),
\end{eqnarray} 
where $\Lambda = \Lambda^a\,T^a$ is the $SU(N)$ valued Lorentz scalar gauge parameter. In addition to the above transformations, the action is also invariant under the vector gauge transformations,
\begin{eqnarray}
\delta_2\, A_{\bar \mu} = 0, \qquad \delta_2\, B_{\bar \mu \bar \nu} = D_{\bar \mu} \Sigma_{\bar \nu} 
- D_{\bar \nu} \Sigma_{\bar \mu}, \qquad \delta_2\, K_{\bar \mu} = - \Sigma_{\bar \mu},
\end{eqnarray} 
where $\Sigma_{\bar \mu} = \Sigma^a_{\bar \mu}\,T^a$ is an $SU(N)$ valued vector.

In order to apply the procedure for the dimensional reduction  as discussed in our previous section, we split the non-Abelian fields as
\begin{eqnarray}
A^a_{\bar \mu}\,T^a &\equiv& A_{\bar \mu} = (A_\mu,\, \varphi),\nonumber\\
B^a_{\bar \mu \bar \nu} \, T^a &\equiv& B_{\bar \mu \bar \nu}  = (B_{\mu\nu},\, \phi_\mu), \nonumber\\
K^a_{\bar \mu}\,T^a &\equiv& K_{\bar \mu} = (K_\mu,\, \rho), \qquad \qquad (\mu, \nu, \cdots = 0,1,2).
\end{eqnarray}
We again assume that all the fields are independent of the third spatial dimension $x^3$.  
Then from Eq.~(\ref{L_NA})  we get the reduced Lagrangian density 
\begin{align}
{\cal L} = &- \frac{1}{4}\, F_{\mu \nu} \cdot F^{ \mu \nu} 
- \frac{1}{4}\, \big[G_{\mu\nu} + g (F_{\mu\nu} \times \rho) \big] \cdot \big[G^{\mu\nu} + g (F^{\mu\nu} \times \rho)\big]
+ \frac{m}{2}\,\varepsilon^{\mu\nu\eta}\,F_{\mu\nu}\cdot \phi_\eta\nonumber\\
&+  \frac{1}{12}\,H_{\mu\nu\eta}\cdot H^{\mu\nu\eta} + \frac{1}{2}\,D_\mu \varphi\cdot D^\mu \varphi  
- \frac{m}{3!}\,\varepsilon^{\mu\nu\eta}\,\varphi\cdot H_{\mu\nu\eta}
+ \frac{g}{2}\,m\,\varepsilon^{\mu\nu\eta}\, \varphi \cdot (F_{\mu\nu} \times K_\eta)\nonumber\\
&+ \frac{g}{2}\, \big[G_{\mu\nu} + g (F_{\mu\nu} \times \rho)\big] \cdot \big[(D^{[\mu}\varphi \times K^{\nu]}) - (\varphi \times B^{\mu\nu})\big]\nonumber\\
&-\frac{g^2}{4}\,\big[(D_{[\mu}\varphi \times K_{\nu]}) - (\varphi \times B_{\mu\nu})\big] \cdot  \big[(D^{[\mu}\varphi \times K^{\nu]})
- (\varphi \times B^{\mu\nu})\big],
\label{L_NAdr}
\end{align}
where we have now defined $G^a_{\mu\nu}\, T^a \equiv G_{\mu\nu} = D_\mu \phi_\nu - D_\nu \phi_\mu\,.$ It is clear from the above Lagrangian density that the scalar field $\varphi$ remains coupled to the  rest of the fields. This happens because of the non-Abelian nature of the theory.

The $SU(N)$ gauge transformations for the three-dimensional fields are
\begin{eqnarray}
&&\delta_1\, A_\mu = D_\mu \Lambda, \quad \delta_1\, \varphi = g(\varphi \times \Lambda), 
\quad \delta_1\, B_{\mu\nu} = g (B_{\mu\nu} \times \Lambda), \quad \delta_1\, \phi_\mu = g (\phi_\mu \times \Lambda)\,,\nonumber\\
&& \delta_1\, K_\mu = g (K_\mu \times \Lambda), \quad \delta_1\, \rho = g (\rho \times \Lambda),
\end{eqnarray}
and the vector gauge transformations reduce to
\begin{eqnarray}
&&\delta_2\, B_{\mu \nu} = D_\mu \Sigma_\nu - D_\nu \Sigma_\mu, 
\quad  \delta_2 \,\phi_\mu = D_\mu \Sigma - g(\varphi \times \Sigma_\mu), \quad \delta_2\, K_\mu = - \Sigma_\mu, \nonumber\\
&& \delta_2\, \rho = - \Sigma, \quad \delta_2 \,A_\mu = 0, \quad \delta_2 \,\varphi = 0.
\end{eqnarray}
The reduced Lagrangian density of Eq.~(\ref{L_NAdr}) respects the above gauge transformations as one can explicitly check that under the two gauge transformations the Lagrangian density transforms by a total derivative, 
\begin{eqnarray}
\delta_1\,{\cal L} = 0, \qquad 
\delta_2\,{\cal L} = \partial_\mu \Big(\frac{m}{2}\,\varepsilon^{\mu\nu\eta}\,F_{\mu\nu}\cdot \phi_\eta\Big). 
\end{eqnarray}
As a consequence, the action integral $S = \int d^3x {\cal L}$ remains invariant.

Earlier when discussing the dimensional reduction of the Abelian theory in Sec.~\ref{topomass}, we defined the Hodge dual of the 3-form field strength $H$ in terms of a scalar field and then integrated out the scalar field. But in the case of the non-Abelian theory, both $H_{\mu\nu\lambda}$ and $B_{\mu\nu}$ remain in the reduced Lagrangian in Eq.~(\ref{L_NAdr}). So in this case we cannot integrate out the scalar field. However, because this is a gauge theory, we can consider it in a particular gauge, namely the axial gauge $A^a_3 = 0\,$ in the four-dimensional theory. In the reduced picture this corresponds to setting $\varphi^a = 0\,.$ As a result, all but the first four terms disappear from the dimensional reduced Lagrangian of Eq.~(\ref{L_NAdr}). The remaining terms of the three-dimensional Lagrangian are  
\begin{align}
{\cal L} =& - \frac{1}{4}\, F_{\mu \nu} \cdot F^{ \mu \nu} 
- \frac{1}{4}\, \big[G_{\mu\nu} + g (F_{\mu\nu} \times \rho) \big] \cdot \big[G^{\mu\nu} + g (F^{\mu\nu} \times \rho)\big]
+ \frac{m}{2}\,\varepsilon^{\mu\nu\eta}\,F_{\mu\nu}\cdot \phi_\eta\nonumber\\
&+ \frac{1}{12}\,H_{\mu\nu\eta}\cdot H^{\mu\nu\eta}, 
\label{L_NAdr2}
\end{align}
and we can see that the first three terms correspond to the Jackiw-Pi model.

The last term in Eq.~(\ref{L_NAdr2})\,, which spoils the equivalence, does not contribute on shell, as we shall see now. First note that the equation of motion for $B_{\mu\nu}$ as derived from Eq.~(\ref{L_NAdr2}) is $(D_\mu H^{\mu\nu\eta})^a = 0\,.$ Defining 
\begin{equation}
f^a = \frac{1}{3!}\,\varepsilon^{\mu\nu\eta}\,H^a_{\mu\nu\eta},
\end{equation}
as the Hodge dual of $H^a_{\mu\nu\eta}\,$ we can write this equation as 
\begin{equation}
(D_\mu f)^a = 0,
\end{equation}
which shows that $f$ is a covariantly constant adjoint scalar. Given a chosen point $x_0$ and a parametrized curve $\gamma(t)\,$ from $x_0$ to some point $x\,,$ we can write a solution to this linear partial differential equation as the path ordered exponential 
\begin{eqnarray}
f(x) = {\cal P}\left[\exp\left(ig\int_{\gamma} dt\, \frac{dz^\mu(t)}{dt}\, A_\mu (z(t))\, \right) \right] f(x_0).
\end{eqnarray}  
The value of $f$ at $x$ thus depends on the starting point $x_0,$ and more importantly, on the chosen path $\gamma$ connecting $x$ with $x_0,$ for any non-flat gauge connection $A_\mu.$

On the other hand, duality implies that we can write 
\begin{equation}
H^a_{\mu\nu\lambda} = \varepsilon_{\mu\nu\lambda} f^a,
\end{equation}
which means $H$ itself, as well as 
\begin{equation}
\frac{1}{12}\,H_{\mu\nu\eta} \cdot H^{\mu\nu\eta} = \frac{1}{2}\,f\cdot f,
\end{equation}
are ill-defined for a generic $A_\mu$, as they are defined in terms of arbitrarily chosen curves connecting the point $x$ with some arbitrarily chosen point $x_0.$ But $H$ is already in the theory as a well-defined local object. Therefore the only allowed value of $H,$ when it satisfies the equation of motion, is $H = 0$.

Thus, topologically massive gauge theory, when dimensionally reduced from four to three dimensions, agrees with the Jackiw-Pi model for both Abelian and non-Abelian cases. It is not difficult to see that the symmetries of the four-dimensional model also reduce correctly to the symmetries of the lower dimensional model in both cases. Detailed calculation shows that the BRST extensions of the two sets of models are also correctly related on shell, i.e. the ghost sector of the four-dimensional model also reduces correctly to the ghost sector of the three-dimensional model. We leave the demonstration of that equivalence for elsewhere.



\begin{thebibliography}{99}
\bibitem{aur} A. Aurilia and Y. Takahashi, {\it Prog. Theor. Phys.} {\bf 66}, 693 (1981).
\bibitem{pk}  D. Z. Freedman and P. K. Townsend, {\it Nucl. Phys.} B {\bf 177}, 282 (1981).
\bibitem{trg} T. R. Govindarajan, {\it J. Phys. G Nucl. Phys.} {\bf 8}, L17 (1982).
\bibitem{a1}  T. J. Allen, M. J. Bowick and A. Lahiri, {\it Mod. Phys. Lett.} A {\bf 6}, 559 (1991).
\bibitem{ru} H. Ruegg and  M. Ruiz-Altaba, {\it Int. J. Mod. Phys.} A {\bf 19}, 3265 (2004).
\bibitem{a2}  A. Lahiri, {\it Phys. Rev.} D {\bf 63}, 105002 (2001). 
\bibitem{a3}  E. Harikumar, A. Lahiri and M. Sivakumar, {\it Phys. Rev.} D {\bf 63}, 105020 (2001). 
\bibitem{a4}  A. Lahiri,  {\it Mod. Phys. Lett.} A {\bf 8}, 2403 (1993). 
\bibitem{a5}  A. Lahiri {\it Phys. Rev.} D {\bf 55} 5045 (1997).
\bibitem{r1}  S. Gupta, R. Kumar and R. P. Malik, {\it Eur. Phys. J.} C {\bf 70}, 491 (2010). 
\bibitem{r2}  R. Kumar and R. P. Malik, {\it Eur. Phys. J.} C {\bf 71}, 1710 (2011). 
\bibitem{r3}  R. Kumar and R. P. Malik, {\it Euro. Phys. Lett.} {\bf 94}, 11001 (2011). 
\bibitem{mieg} J. Thierry-Mieg and L. Baulieu, {\it Nucl. Phys.} B {\bf 228} 259 (1983). 
\bibitem{lee} D. S. Hwang and C. -Y. Lee, {\it J. Math. Phys.} {\bf 38}, 30 (1997).
\bibitem{a6}  A. Lahiri and D. Mukhopadhyay, {\it Phys. Rev.} D {\bf  90}, 025015 (2014).
\bibitem{hen} M. Henneaux, V.E.R. Lemes, C.A.G. Sasaki, S.P. Sorella, O.S. Ventura and L.C.Q. Vilar, {\it Phys. lett.} B {\bf 410}, 195 (1997).
\bibitem{j1}  S. Deser, R. Jackiw and S. Templeton, {\it Phys. Rev. Lett.} {\bf 48}, 975 (1982).
\bibitem{j2}  S. Deser, R. Jackiw and S. Templeton, {\it Ann. Phys.} (NY) {\bf 140}, 372 (1982).
\bibitem{j3}  R. Jackiw and S-Y. Pi.,  {\it Phys. Lett.} B {\bf 403}, 297 (1997).
\bibitem{day} \"O. F. Dayi, {\it Mod. Phys. Lett. A} {\bf 13}, 1969 (1998). 
\bibitem{cim} O. M. Del Cima, {\it J. Phys. A: Math. Theor.} {\bf 44}, 352001 (2011). 
\bibitem{r4}  S. Gupta, R. Kumar and R. P. Malik, {\it Can. J. Phys.} {\bf 92}, 1033 (2014). 
\bibitem{r5}  S. Gupta and R. Kumar, arXiv:1411.6357[hep-th].
\bibitem{sch} J. Scherk and J. H. Schwarz, {\it Phys. Lett.} B {\bf 57}, 463 (1975). 
\bibitem{c1}  E. Cremmer and J. Scherk, {\it Nucl. Phys.} B {\bf 103}, 399 (1976).
\bibitem{c2}  E. Cremmer and J. Scherk, {\it Nucl. Phys.} B {\bf 108}, 409 (1976).
\bibitem{sal} A. salam and J. Strathdee, {\it Ann. Phys.} (NY) {\bf 141}, 316 (1982). 
\bibitem{siv3} T. R. Govindarajan, S. D. Rindani and M. Sivakumar, {Phys. Rev.} D {\bf 32}, 454 (1985).
\bibitem{net1} H. Belich Jr., M. M. Ferreira Jr., J. A. Helay\"el-Neto and M. T. D. Orlando, {\it Phys. Rev.} D {\bf 67}, 125011 (2003)
               [Erratum-ibid. D {\bf 69}, 109903 (2004)].
\bibitem{sch1} J. Schwinger, {\it Phys. Rev.} {\bf 125}, 397 (1962).
\bibitem{sch2}  J. Schwinger, {\it Phys. Rev.} {\bf 128}, 2425 (1962).            
\end{thebibliography}
\end{document}